\documentclass[a4paper]{aa}

\usepackage{graphicx}
\usepackage{natbib}

\usepackage{txfonts}

\newcommand{\smin}{\Sigma_{s,min}}

\begin{document}
\titlerunning{Giant planets in disks with different metallicities}
\title{Formation of giant planets in disks with different metallicities}
\author{K. Kornet\inst{1}\thanks{Currently at MPIA, K\"onigstuhl 17,  69117 Heidelberg}  P. Bodenheimer\inst{2}, M. R\'o\.zyczka\inst{1}, 
T.F. Stepinski\inst{3}}
\institute{Nicolaus Copernicus Astronomical Center , Bartycka 18 , Warsaw, 
PL-00-716, Poland\\ \email{kornet@camk.edu.pl; mnr@camk.edu.pl}
\and
UCO/Lick Observatory, Department of Astronomy and Astrophysics, University of
California, Santa Cruz, CA 95064, USA\\ \email{peter@ucolick.org}
\and
Lunar and Planetary Institute, 3600 Bay Area Blvd., Houston, TX
77058, USA\\ \email{tom@lpi.usra.edu}
}

\abstract{We present the first results from simulations of processes
leading to planet formation in protoplanetary disks with different 
metallicities. For a given metallicity, we construct a two-dimensional
grid of disk models with different initial masses and radii ($M_0$, $R_0$). 
For each disk, we follow the evolution of gas and solids from an early 
evolutionary stage, when all solids are in the form of small dust grains, 
to the stage when most solids have condensed into planetesimals. Then, 
based on the core accretion - gas capture scenario, we estimate the 
planet-bearing capability of the environment defined by the final 
planetesimal swarm and the still evolving gaseous component of the disk. 
We define the probability of planet-formation, $P_p$, as the normalized
fractional area in the ($M_0$, $\log R_0$) plane populated by disks that 
have formed planets inside 5~AU. 
With such a definition, and under the assumption that the population
of planets discovered at $R$ $<$ 5~AU is not significantly contaminated by 
planets that have migrated from $R$ $>$ 5~AU, our results agree fairly well 
with the observed dependence between the probability that a star harbors 
a planet and the star's metal content. The agreement holds for the disk 
viscosity parameter $\alpha$ ranging from $10^{-3}$ to $10^{-2}$, and it
becomes much poorer when the redistribution of solids relative to the gas 
is not allowed for during the evolution of model disks.}

\maketitle

\section{Introduction}

Well over a hundred extrasolar planets have already been catalogued.
It is very likely that radial velocity surveys performed in the solar
neighbourhood have discovered a significant fraction of all Jupiter-like 
planets with periods shorter than $\sim$10 yr (or, equivalently, with 
orbital radii smaller than $\sim$5~AU), whose parent stars exhibit periodic 
radial velocity variations with amplitudes $\geq 10$~m~s$^{-1}$ Such a
collection of planets provides a good data set with which predictions of
different theories of planet formation can be compared \citep[e.g.][]{ida04}.

\begin{figure}
\resizebox{\hsize}{!}{\includegraphics{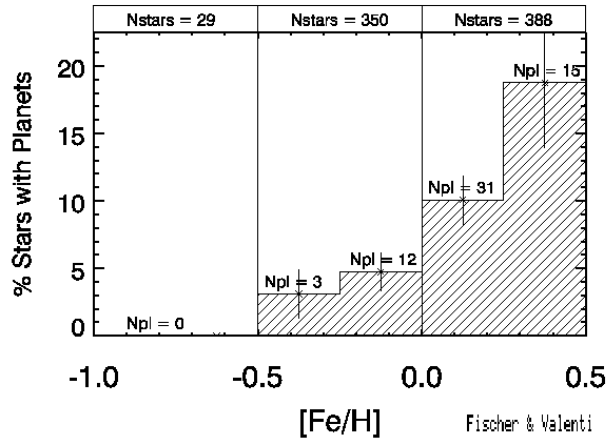}}
\caption{The probability of planet occurrence as a function of the star's 
  iron abundance (on the standard logarithmic scale with zero representing 
  the solar value). Reproduced with permission from \cite{debra03}}
\label{f:fe_bar}
\end{figure}

Current observations strongly suggest that  planet-bearing stars
tend to have higher metallicities than field stars (Fig.
\ref{f:fe_bar}). This correlation was reported for the first time by
\cite{santos00}, and later confirmed by \cite{debra03}. The effect
finds a natural explanation in the core accretion - gas capture (CAGC)
model for giant planet formation. The model predicts a slow accretion
of planetesimals onto a protoplanetary core until the mass of the core
reaches a few $M_\oplus$. Then, gas accretion becomes significant, and
around the core an extended envelope is formed, whose mass increases
faster than that of the core. Eventually, the envelope becomes more
massive than the core, and a runaway accretion of gas ensues, which
terminates when tidal effects become important or the protoplanetary
disk is dissipated. The length of the core accretion phase decreases
with increasing surface density of the planetesimal swarm
\citep{pollack96}, which, in turn, is an increasing function of the original metal content of the protoplanetary disk and its
central star. Consequently, the giant planets are expected to form
more easily in disks with higher metallicities.

However, dust evolves due to different physical processes than  the
gaseous component of the disk \citep{weiden93} As a result,
a significant redistribution of solids takes place with respect to the 
gas, and, locally, the surface density of solids can be considerably enhanced 
compared to the initial one \citep{weiden03,SV97}. Thus,
statistical tests of the CAGC scenario must be based on evolving models of 
protoplanetary disks rather than static configurations which only serve as
mass supply to planetary cores and envelopes. Such models must include 
the global evolution of solids. An extensive survey of such models was 
reported by \cite{kac1}, and applied to the problem of extrasolar 
planets formation by \cite{kac2}. In the latter paper it was found 
that effects of dust evolution allow for \textit{in situ} formation of giant 
planets as close as 2 AU from the star. Here we obtain a much larger sample of 
disk models with different primordial metallicities, estimate how many of them
could form planets, and compare the estimates to the observed correlation
between stellar iron abundance and the probability of planet occurrence. 

In \S 2 we briefly explain our approach to the evolution of a protoplanetary 
disk and planet formation. The results are presented in \S 3, while in \S 4 
we summarize them and critically assess the simplifications on which our 
method is based.

\section{Method of calculation}
\subsection{The disk}
\label{sect: disk}

We model the protoplanetary disk as a two component fluid, consisting
of gas and solids. The gaseous component is described by the 
analytical model of \cite{s98} which gives the surface
density of gas, $\Sigma_g$, as a function of radius $r$ and time $t$,
in terms of a solution to the viscous diffusion equation. The viscosity
coefficient is given by the standard $\alpha$ prescription. All other
quantities characterizing the gas are calculated from a thin disk
approximation, assuming vertical thermal balance. To take into account
the effect of different metallicities on the structure of the gaseous
disk, we scaled the opacity in Stepinski's model by a constant 
factor $Z$, independent of $r$ or $t$ and representing the metallicity 
of the disk expressed in solar units. Such a simple-minded approach is 
justified by the fact that the opacity in protoplanetary disks is mainly due
to dust grains and molecules.

The main assumptions underlying our approach to the evolution of
solids are (1) at each radius the particles have the same size (which
in general varies over time), (2) there is only one component of dust,
in this case corresponding to high-temperature silicates with the
evaporation temperature $T_{evap}$ = 1350 K and bulk density 6.3 g
cm$^{-3}$ (this choice is justified by the fact that in most cases the
surface density of a water-ice planetesimal swarm would be too low in the
range of distances considered here, i.e. $r\le5$ AU, to enable 
formation of a giant planet in a reasonable time; see also
\cite{kac4}), (3) all collisions between particles lead to
coagulation, (4) when the temperature exceeds $T_{evap}$, local solids
immediately sublimate and the vapor evolves at the same radial
velocity as the gas component, (5) when the disk temperature falls
below $T_{evap}$, local vapor immediately condenses into grains with a
radius of 10$^{-3}$ cm (minimum allowed by the numerical code), (6)
the radial velocity of solid particles is entirely determined by the
effects of gas drag, (7) the relative velocities of solid particles at
which they collide are computed according to the turbulent model
described by \cite{SV97}, (8) the evolution of solids does not affect
the evolution of gas. At each radius, the vertical extent of the solid
particle distribution is calculated and is evolved in time, so the
effect of sedimentation of solids toward the midplane of the disk is
taken into account.  All assumptions and approximations are discussed
in \cite{SV97}, \cite{s98} and \cite{kac1}.

The evolution of solids is governed by two equations. The first of
them is the continuity equation for surface density of solid material,
$\Sigma_s$. The second one, describing the evolution of grain sizes,
can be interpreted as the continuity equation for size-weighted
surface density of solids, $\Sigma_a\equiv a(r)\Sigma_s$, where $a(r)$
is the radius of solid particles at a distance $r$ from the star. The
equations are solved numerically on a moving grid whose outer edge
follows the outer edge of the dust disk. The details of the method
can be found in \cite{kac1}.

We parameterize the initial conditions by the initial mass of the disk,
$M_0$ (in $M_\odot$) and the decimal logarithm of the initial outer
radius of the disk, $R_0$ (in AU). From these two
quantities the initial total angular momentum of the disk $j_0$, which
we used in our previous calculations, can be calculated from the
formula
\begin{equation}
  j_0=6.9 M_0 R_0^{1/2} M_\star^{1/2}
  \label{eq: jdef}
\end{equation}
in units of $10^{52}$ g cm$^2$ s$^{-1}$, where $M_\star$ is the mass
of the central star expressed in $M_\odot$ (see also eq. (21)in
\cite{s98}.  In the present survey $M_\star=1$. Once $M_0$ and $R_0$
are specified, the initial distribution of gas surface
density $\Sigma_{g,0}\equiv\Sigma_g(r,t=0)$, as well as the
complete evolution  of the gaseous disk, are obtained from the
analytical model of \cite{s98}.

We assume that at $t=0$ all solids are in the form of grains with
radii $a=10^{-3}$~cm, well mixed with the gas. Thus, the ratio of
$\Sigma_{s,0}$ to $\Sigma_{g,0}$ is independent of $r$, and proportional to
disk metallicity, $Z$. For a given metallicity, the initial surface
density distribution of solids is
\begin{equation}
   \Sigma_{s,0}(r)= 6\times 10^{-3} Z \Sigma_{g,0},
   \label{eq: sigmainit}
\end{equation}
where $Z$ is expressed in solar units. 

\subsection{The planets}
\label{sect: planets}

We model the formation of a giant planet \textit{in situ}, i.e. we do
not include effects of migration, and the orbital parameters of the
planet do not vary in time.  Its evolution is followed according to
the approach developed by \cite{pollack96} and \cite{boden00}. The
planet consists of a solid core and a gaseous envelope, both of which
accrete mass from the disk. Our implementation of the evolutionary
procedure is based on the following assumptions: (1) core accretion
starts when all solids are in the form of planetesimals with the same
radius $a$ = 2 km, (2) the surface density of solids in the feeding
zone decreases with time as material accretes onto the core, (3) at
each time the planetesimals are well mixed through the feeding zone of
the planet; thus $\Sigma_s$ is always uniform in space, but usually
decreasing with time, (4) the planetesimals do not migrate into the
feeding zone from outside, or vice versa, but they can be overtaken by
the boundary of the feeding zone as it expands due to the growing mass
of the planet.

With these assumptions, the minimum surface density of the
planetesimal swarm $\smin$, needed to form a Jupiter-mass planet in
less than 3~Myr can be determined as a function of distance $r$ from
the star. The details of the procedure are discussed in \cite{kac2};
here we only note that 2-km planetesimals are on essentially Keplerian
orbits, i.e. drag force has negligible effects on their distribution
within the gaseous disk. The limit of 3~Myr is supported by recent
observations \citep[e.g.][]{sicilia04, haisch}, and it leaves
a reasonable safety margin for the planet to complete its evolution
before the canonical time of disk dispersal ($10^7$ yr) elapses.

\section{Results}

To cover the range of masses and sizes of protoplanetary disks
indicated by observations, we consider 11 values of $M_0$ uniformly
distributed between 0.02 and 0.2, and 10 values of $R_0$ so chosen
that $\lg R_0$ is uniformly distributed between 1 and 4. We evolve
this set of models for seven different metallicities ($Z$ = 0.2, 0.5,
0.75, 1, 1.5, 2.0, 2.5 and 3.0), and three values of the viscosity
coefficient ($\alpha$ = $10^{-3}$, $10^{-2}$ and $10^{-1}$). Using the
procedure described in \S \ref{sect: disk}, we follow each model
until all solids are in the form of "Keplerian" planetesimals or they
have accreted onto the star. Next, each model with planetesimals is
scanned to determine whether anywhere in the disk the surface density
of the final planetesimal swarm, $\Sigma_{s,f}(r)$ exceeds $\smin(r)$.
Models with a range of $r$ such that $\Sigma_{s,f}(r)\ge\smin(r)$ are
{\it planet bearing}; however for further analysis only those that
form planets closer than 5~AU from the star are selected. We
introduce this limitation in order to account for the fact that (as we
already mentioned in \S 1) the sample of observed extrasolar planets
is reasonably complete to $\sim$5~AU only.

For given $\alpha$ and $Z$ our grid of $M_0$ and $\lg R_0$ may be
thought of as a {\it nonrandom} sample drawn from the {\it unknown}
ensemble of disk masses and radii. Let ${\cal R}_{p,5}$ be the region
in the $[M_0, \lg R_0]$ plane occupied by models that form planets at
$r\le5$~AU from the star, and let $A_{p,5}$ be the area of ${\cal
  R}_{p,5}$. The rate of planet occurrence {\it within the sample} is
given by the ratio
\begin{equation}
  P_p=\frac{A_{p,5}}{A},
  \label{eq: percentage}
\end{equation}
where $A$ is the area of the region occupied by the whole grid of  models.

\begin{figure}
\resizebox{\hsize}{!}{\includegraphics[angle=-90]{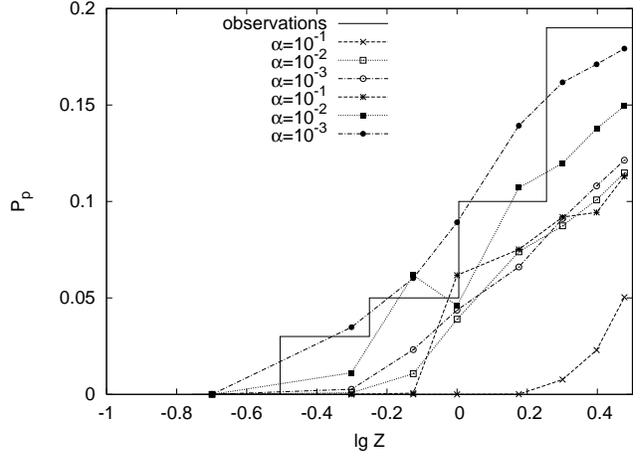}}
\caption{The rate of planet occurrence as a function of the primordial
  metallicity of protoplanetary disks. Lines marked with different symbols 
  are obtained for different values of $\alpha$, as labeled in the upper left
  corner. Filled symbols: models with the redistribution of solids; open symbols:
  models with a constant $\Sigma_s/\Sigma_g$ ratio. The histogram shows the 
  observational data.}
\label{f:rout_bezw}
\end{figure}

The results are presented in Fig. \ref{f:rout_bezw}. It shows the rate
of planet occurrence as a function of disk metallicity for different
values of $\alpha$.  As one might expect, $P_p$ is an increasing
function of $Z$: in more metal-rich disks there is more solid
material, so the planets form more easily. We believe that deviations
from that rule result from the poor resolution of our grid of models.

\begin{figure}
\resizebox{\hsize}{!}{\includegraphics[]{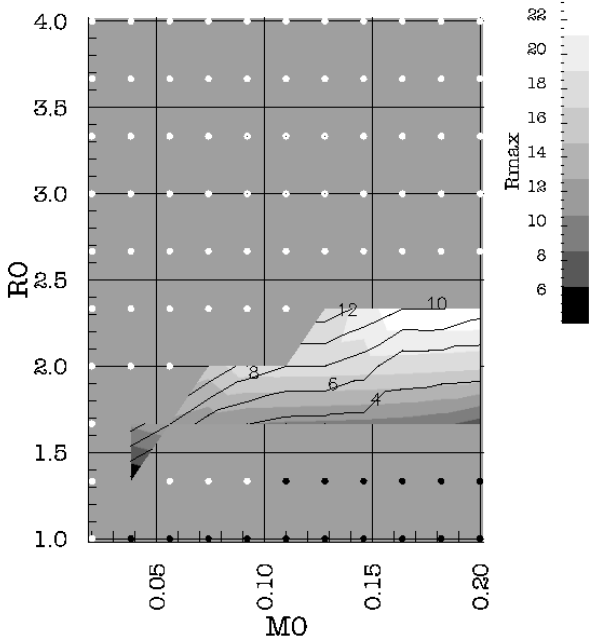}}
\caption{ As functions of the initial disk mass $M_0$ (in solar
  masses) and the logarithm of the initial radius $R_0$ in
  astronomical units, the contours and grey scale, respectively, give
  the inner and outer radius of the region around the central star
  where giant planet formation is possible in a maximum of 3 Myr.
  White circles indicate disk models where the solid surface density
  is everywhere below the critical value for planet formation. Black
  circles indicate disks in which all of the solid material accretes
  onto the star. The disk viscosity parameter is $\alpha = 1 \times
  10^{-2}$, and the primordial metallicity is solar.}
\label{fig:grid}
\end{figure}

There is also a monotonic trend seen when $P_p$ is considered as a
function of $\alpha$ at constant $Z$: giant planets form more easily in
less viscous disks. This is because for every $\alpha$ and every $Z$
the lower border of the planet-bearing region runs very close, and nearly
parallel, to the line ${\cal L}$ that separates models in which all
solids accrete onto the star from those in which some solids survive
(see Fig. \ref{fig:grid}). The observed trend results from two
competing factors. On the one hand, to a good approximation ${\cal L}$
marks the location of models in which the initial temperature at the
outer edge of the disk is equal to $T_{evap}$ \citep{kac1,kac2}. Among
disks with the same $M_0$ and $R_0$, those with smaller $\alpha$ are
cooler as a result of lower energy generation rate.  Thus, when
$\alpha$ decreases, ${\cal L}$ must move toward smaller values of
$R_0$, and the area of the planet-bearing region increases. On the
other hand, decreasing $\alpha$ reduces the maximum $R_0$ for which
planet formation is possible in disks with the same $M_0$. This is
because in cooler disks the solids gain smaller inward velocities
\citep{weiden77,kac4}. As a result, for the same disk parameters
except for a lower $\alpha$, the redistribution of solids is less
efficient, and the final planetesimal swarm is more extended. A more
extended swarm implies lower surface density of solids and less
favorable conditions for planet formation. In particular, when we move
on the $[M_0, \lg R_0]$ plane along an $M_0$ = const line such that
$R_0$ increases, we leave the planet-bearing region sooner (in other
words, decreasing $\alpha$ causes the upper boundary of the
planet-bearing region to move downwards on the $[M_0,R_0]$ plane). Our
calculations show that the first of these two effects is the dominant
one. Therefore, for a given metallicity the fraction of disks with
giant planets increases with decreasing $\alpha$.
 
To illustrate the importance of the redistribution of solids for the
process of planet formation, we calculated $P_p$ for the same grid 
of initial conditions, but based on $\Sigma_{s,0}$ instead of 
$\Sigma_{s,f}$. The results are shown in Fig. \ref{f:rout_bezw}.  
As expected, the redistribution of solid material is most important for 
$\alpha=10^{-1}$, but even for $\alpha=10^{-3}$ it cannot be neglected. 
For example, with $\alpha=10^{-3}$ and $Z=3$, including redistribution  
increases the ratio of planet occurrence by a factor $\approx 1.5$. 
For lower metallicities the effect is even stronger. In particular, it 
causes $Z_{min}$, the minimum  $Z$ at which planets can form, to decrease 
from$\sim 0.7$ to $\sim 0.4$ and from $\sim 1.2$ to $\sim 0.85$, 
respectively, for $\alpha$ = $10^{-3}$ and $10^{-1}$. It can be seen that 
even in low-metallicity disks $\Sigma_s$ can locally grow above $\smin$, 
creating conditions favorable for planet formation.

\begin{figure}
\resizebox{\hsize}{!}{\includegraphics[angle=-90]{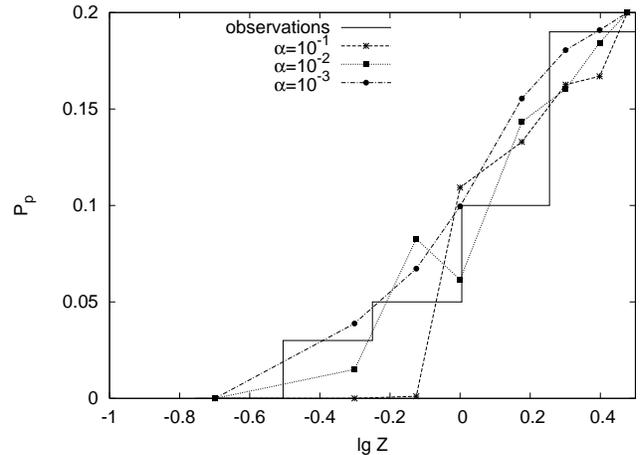}}
\caption{Rate of planet occurrence as a function of the primordial
  metallicity in models with the redistribution of solids included. 
  Same as in Fig. \ref{f:rout_bezw}, but all curves are normalized 
  to match the observational data for $\lg Z$ = 0.4.} 
\label{f:rout_sc}
\end{figure}

\begin{figure}
\resizebox{\hsize}{!}{\includegraphics[angle=-90]{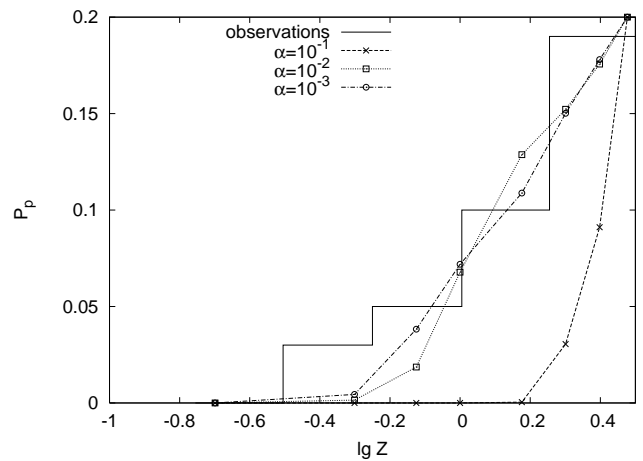}}
\caption{Rate of planet occurrence as a function of the primordial
  metallicity in models without the redistribution of solids. 
  Same as in Fig. \ref{f:rout_bezw}, but all curves are normalized 
  to match the observational data for $\lg Z$ = 0.4.}
\label{f:rout_orig}
\end{figure}

With $\alpha=10^{-3}$ and the redistribution of solids included, the
calculated rate of occurrence of planets is in good agreement with
observational data. Obviously, it is a mere coincidence, without any
practical meaning. Our rates strongly depend on the extent and
position of the grid of initial conditions in the $[M_0,R_0]$ plane.
Suppose that we increase the maximum $\lg R_0$ above the considered
value of 4. The rate of planet occurrence would decrease, because we
would add only models in which $\Sigma_{s,f}$ is not high enough to
satisfy the requirements for planet formation. To make our results
independent of this effect, for every value of $\alpha$ we normalized
$P_p(Z)$ in such a way as to reproduce the observational value of
$\sim 0.2$ for $Z\approx3$. The normalized rates are shown in
Fig. \ref{f:rout_sc}. Now, curves for both $\alpha=10^{-3}$ and
$10^{-2}$ fall in the range of observational values, while disks with
$\alpha=10^{-1}$ still produce too few planets for small values of
$Z$. For comparison, in Fig. \ref{f:rout_orig} we shows the normalized
rates obtained from models without the redistribution of solids. While
for $\alpha=10^{-3}$ and $10^{-2}$ they are consistent with the
observational data for metal rich disks, they are significantly lower
for small values of $Z$, and they give too high values of
$Z_{min}$. For $\alpha=10^{-1}$, the results do not match observations
except for $Z\sim3$, where the agreement is enforced by the
normalization procedure.

\section{Discussion and conclusions }

Based on a simple approach to the evolution of solids in protoplanetary 
disks we calculated rates of giant planet occurrence on orbits smaller 
than 5~AU. We find that in more metal-rich disks the planets form more 
easily, and we qualitatively reproduce the observed correlation between 
the probability that a giant planet would be found at a star and the 
metal content of that star. We are able to match this correlation 
quantitatively for disk models with the viscosity parameter $\alpha$ 
ranging from $\sim 10^{-3}$ to $\sim10^{-2}$. We also show that in 
the low-metallicity regime the calculated occurrence rates agree 
with the observed ones only if the redistribution of solids is included 
in the models. In such models the surface density of solids locked in 
the final planetesimal swarm can be locally much higher than the initial 
surface density of the dust. This effect is particularly important if 
we want to match the observed value of limiting stellar metallicity 
$Z_{min}$ below which no planets have been detected. Again, only models 
which allow for the redistribution of solids are successful, while 
models which do not include this effect produce $Z_{min}$ which is 
significantly too high. This finding supports our approach to the 
evolution of protoplanetary disks, whose key ingredient is the 
radial drift of solids due to the gas drag. 

\begin{figure}
\resizebox{\hsize}{!}{\includegraphics[angle=-90]{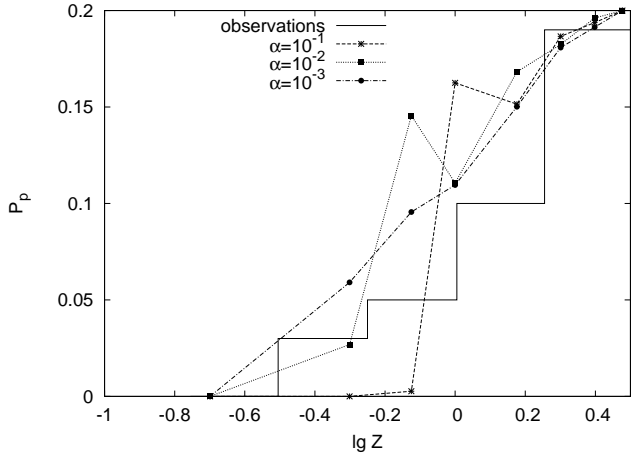}}
\caption{Same as Fig. \ref{f:rout_sc}, but based on disk models with initial
 parameters uniformly distributed on the $[M_0, \lg j_0]$ plane.}
\label{f:j_sc}
\end{figure}

The above conclusions are based on the assumption that our grid of models 
is a representative sample of the real distribution ${\cal P}(M_0,R_0)$ 
of initial masses and radii of protoplanetary disks, which is not known. 
Such an assumption is acceptable provided that the real disks are uniformly 
(or almost uniformly) distributed on the $[M_0,\lg R_0]$ plane. Adopting 
it we demand that very large disks be relatively rare, which seems to be 
an entirely reasonable requirement (note that ${\cal P}(M_0,R_0)$ {\it must} 
approach 0 for $R_0\rightarrow\infty$). We tested the sensitivity of our 
results to the way the initial parameters are sampled by repeating the 
calculations with initial models uniformly distributed on the $[M_0,\lg j_0]$ 
plane. The results, normalized to the observational data the same way as 
before, are shown in Fig. \ref{f:j_sc}. The agreement is poorer, but the 
same general trend is still well visible. Moreover, we see that $Z_{min}$ 
does not depend on the sampling procedure (this conclusion is restricted to 
samples which do not include extremely massive disks with $M_0>0.2~M_\odot$; 
note however that such disks would most likely be gravitationally unstable). 

Our results are based on a highly simplified scenario of the evolution 
of solids in protoplanetary disks. The assumption concerning the size 
distribution of solid particles, according to which at each distance 
from the star all particles have the same diameter, seems to be 
particularly radical. However, detailed work by \cite{morfill85},\cite{mizuno88} and \cite{weiden97} showed that the 
distribution of solids quickly converges to a stage in which most of 
the mass is concentrated in a narrow range of sizes at the momentary 
maximum size. Their conclusion has recently been corroborated by the 
behaviour of solids in the two-dimensional disk model by 
\cite{weiden03}. Thus, despite its simplicity, our 
our approach to the size disitribution function can be regarded as a
reasonable approximation.

Our next radical assumption is the 100\% efficiency of coagulation 
In reality, one hardly expects that collisions between solid particles
always lead to sticking without fragmentation. However, calculations 
reported by \cite{kac4} show that if the main process responsible
for particle growth is collisional coagulation, then its efficiency
{\it must} be high - otherwise the disk would rapidly lose most of its 
solid material, disabling planet formation or severely diminishing its
possibility. Note also that recent observations by \cite{chakrab04} 
indicate that asteroid-sized bodies can exist in disks as young as 0.1 Myr. 
Again, if they have grown due to coagulation, the sticking efficiency must
have been very high. 

Migration of planets due to gravitational interactions with the disk
is not included in our approach. Obviously, we do not claim
that the process of migration does not operate. On the contrary, we
have to assume some form of  redistribution of planets in order to
explain the origin of "hot Jupiters", whose orbits are much tighter
than the tightest orbits allowed by our models ($\sim 1$~AU).
However, as long as we are interested only in global rates of planet
occurrence, and not in the distribution of orbital parameters, our
results remain valid provided that the number of planets which migrate
to $r<5$~AU from larger orbits is negligible compared to the number of
planets which are born there.

The fact that such a simple model can reproduce the available data has
two possible interpretations. According to the pessimistic one we
observe a mere coincidence which does not have any physical meaning.
According to the optimistic one we identify the most important
process governing the evolution of solids in a protoplanetary disk
(radial drift, induced by the gas drag, and associated with
collisional coagulation), and we provide support for the core
accretion - envelope capture scenario of giant planet formation. More
sophisticated disk models based on fewer assumptions, which we are
presently working on, should help us decide which of the two is true.
 
Acknowledgments. KK and MR were supported by the grant No. 1 PO3D 026
26 from the Polish Ministry of Science. They gratefully acknowledge
benefits from the activities of the RTN network "PLANETS" supported by
the European Commission under the agreement No
HPRN-CT-2002-00308. This work was supported in part by the NASA
Origins of Solar Systems program through grant NAG 5-13285 to the
University of California, Santa Cruz.  KK was also supported by the
German Research Foundation (DFG) through the Emmy Noether grant WO
857/2-1. TS was supported by the LPI, which is operated by USRA under
contract CAN-NCC5-679 with NASA. This is LPI contribution No. 1214.

\bibliography{metallicity}
\bibliographystyle{aa}

\end{document}